\documentclass[twocolumn,aps,prl,amssymb,showpacs,relax]{revtex4-1}
\usepackage[dvips,final]{graphicx}
\usepackage{graphicx}
\usepackage{epstopdf}
\begin{document}

\title{Evidence of Josephson-coupled superconducting regions at the interfaces of
Highly Oriented Pyrolytic Graphite}
\author{A. Ballestar}
\author{J. Barzola-Quiquia}
\author{P. Esquinazi}\email{esquin@physik.uni-leipzig.de}
\affiliation{Division of Superconductivity and Magnetism, Institut
f\"ur Experimentelle Physik II, Universit\"{a}t Leipzig,
Linn\'{e}stra{\ss}e 5, D-04103 Leipzig, Germany}

\begin{abstract}

Transport properties of a few hundreds of nanometers thick (in
the graphene plane direction) lamellae  of highly oriented
pyrolytic graphite (HOPG) have been investigated. Current-Voltage
characteristics as well as the temperature dependence of the
 voltage at different fixed input currents provide evidence for Josephson-coupled
superconducting regions embedded in the internal two-dimensional
interfaces, reaching zero resistance at low enough temperatures.
The overall behavior indicates the existence of superconducting
regions with critical temperatures above 100~K at the internal
interfaces of oriented pyrolytic graphite.

\end{abstract}


\maketitle

Superconductivity in doped graphite goes back to 1965 when it was
first observed in the potassium intercalated graphite
$C_8K$\cite{han65}. Since then a considerable amount of studies
reported this phenomenon, reaching critical temperatures $T_c \sim
10$~K in intercalated graphite\cite{wel05,eme05} and above 30~K -
though not percolative - in some HOPG samples\cite{yakovjltp00} as
well as in doped graphite\cite{silvaprl,Kop04,fel09,kopejltp07}.
Theoretical works that deal with superconductivity in graphite as
well as in graphene have been published in recent years. For
example, $p$-type superconductivity has been predicted to occur in
inhomogeneous regions of the graphite structure \cite{gon01} or
$d-$wave high-$T_c$ superconductivity\cite{nan12} based also on
resonance valence bonds\cite{doni07}, or at the graphite surface
region due to a topologically protected flat band\cite{kop11}.
Following a BCS approach in two dimensions (with anisotropy)
critical temperatures $T_c \sim 60~$K have been estimated if the
density of conduction electrons per graphene plane increases to $n
\sim 10^{14}~$cm$^{-2}$, a density that might be induced by
defects and/or hydrogen ad-atoms\cite{garbcs09} or by Li
deposition\cite{pro12}. Further predictions for superconductivity
in graphene support the premise that $ n > 10^{13}~$cm$^{-2}$ in
order to reach $T_c
> 1~$K\cite{uch07,kop08}.

The possibility of high-temperature superconductivity at surfaces
and interfaces has attracted  the attention of the low-temperature
physics community also since the  60's\cite{gin64}. Recently,
superconductivity has been found at the interfaces between oxide
insulators \cite{rey07} as well as between metallic and insulating
copper oxides  with  $T_c \gtrsim 50~$K\cite{goz08}. In case of
doped semiconductors the example of Bi is of interest;  interfaces
in Bi-bicrystals of inclination type show superconductivity up to
21~K, although Bi bulk is not a superconductor\cite{mun08}. These
two independently obtained indications, the possible existence of
high-temperature superconductivity in graphite/graphene and the
special role of interfaces\cite{bar08,sru11}
stimulated us to pursue the study of the transport properties of a
bundle of internal interfaces in bulk HOPG samples.

\begin{figure}
\begin{center}
\includegraphics[width=.95\columnwidth]{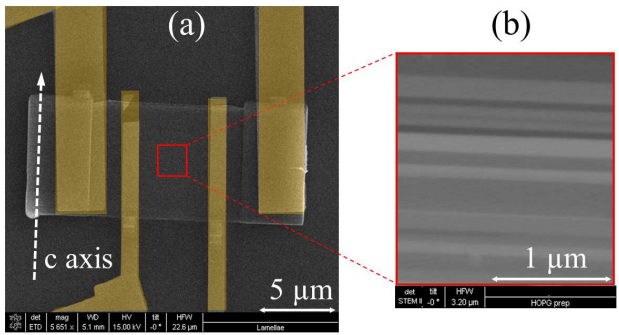}
\caption[]{(a) Scanning Electron Microscopy (SEM) image of sample
L3 on a Si/SiN substrate where the yellowish colored areas are the
electrodes. A four points configuration has been prepared with the
outer electrodes used to apply current and the inner ones to
measure the voltage drop. As shown in the picture, the c-axis runs
parallel to the substrate surface and normal to the current
direction. (b) Transmission Electron Microscopy (TEM) image of one
HOPG lamella. A different brightness corresponds to different
orientation within the $a-b$ plane of the  crystalline regions of
Bernal type with thickness between 30 and 200~nm.} \label{Fig1}
\end{center}
\end{figure}

Transmission electron microscope (TEM) studies on HOPG samples
revealed single crystalline regions of Bernal graphite of
thickness in the $c-$axis direction between 30~nm and $\sim
150~$nm\cite{bar08}. As example, we show in Fig.~\ref{Fig1}(b) a
typical TEM picture of the HOPG samples studied in this work. The
different gray colors shown in Fig.~\ref{Fig1}(b) indicate sightly
different angle misalignments about the c-axis between each other
and the existence of very well defined two-dimensional interfaces
between them. We note that rotations up to
30$^\circ$ between the graphene layers from neighboring graphite
regions has been seen by high resolution TEM in few layers
graphene sheets\cite{war09}. The electrical response of those interfaces is the
aim of our  experimental study.

Thin TEM lamellae  have been prepared from a HOPG (ZYA grade,
0.3$^\circ$ rocking curve) bulk sample using a dual beam
microscope (FEI Nanolab XT200). To avoid contamination and
structural disorder the lamellae were cut after depositing a
protective layer of 300~nm of tungsten carbide using the electron
gun (EBID) on top of the HOPG surface.  The Ga$^+$ ion beam was
used to cut the lamellae of thickness  between $\sim 300$ and
800~nm in the a-b graphene plane direction and lengths up to $\sim
17~\mu$m. After transferring the lamellae to an insulating Si/SiN
substrate, electron beam lithography followed by thermal
evaporation of Pt/Au was used to make the four electrical contacts
allowing us the measurement of the voltage drop of several
interfaces in parallel to the graphene planes along a length
 $\sim 2 \ldots \sim 8~\mu$m. More than ten
lamellae were studied. Although we found qualitatively similar
behavior, the main difference in their characteristic transport
properties depends on the used thickness, i.e. the observed
Josephson-like $I-V$ characteristics and the granular
superconductivity behavior are observed at lower temperatures or
even vanish the smaller the thickness (in the $a,b$ plane) of the
lamella. In this paper we present and discuss the results of four
of them (L1-L4) with thickness ($\sim 500$nm, $\sim 800$nm, $\sim
300$nm, $\sim 800$nm), respectively.

\begin{figure}[]
\begin{center}
\includegraphics[width=.95\columnwidth]{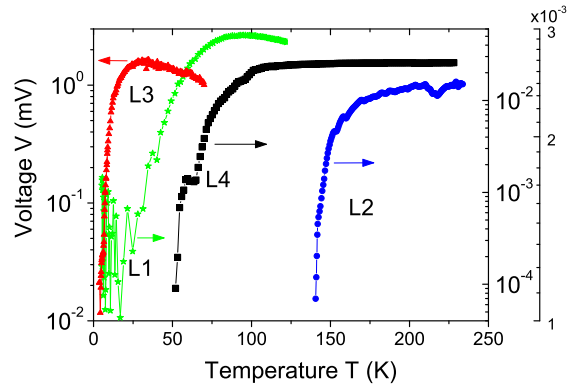}
\caption[]{Temperature dependence of the voltage for four samples
measured with small input currents. A clear drop in the measured
voltage is observed at $15~$K$ < T < 150$~K upon sample. For the
sample L4 the region near the onset of the voltage decrease is
shown (second right $y$-axis).} \label{Fig2}
\end{center}
\end{figure}

Four probe electrode configuration as shown  in
Fig.~\ref{Fig1}(a), and also the van der Pauw configuration
(contacts at lamella edges) were used to measure the temperature
($T$) dependence of the voltage ($V$) at constant input current
($I$) and the $I-V$ characteristic curves. Figure~\ref{Fig2} shows
the $T$-dependence of the measured voltage at $I= 1$~nA (L1,L3)
and 100~nA (L2,L4). A clear drop in the measured voltage is
observed, upon sample at different ``critical" temperatures $T_c$
between $\sim~$15~K to $\sim$~150~K. This $T_c$ reflects the
temperature above which the Josephson coupling between
superconducting patches at some of the interfaces vanishes. At low
enough temperatures and currents zero resistance states were
reached for L1, L3 and L4. Negative saturation voltages (L2)
instead of zero are obtained
 for the van der Pauw configuration, which can be quantitatively explained using a simple  Wheatstone
bridge circuit and assuming two Josephson junctions with different
$T_c$'s as explained below.

\begin{figure*}[]
\begin{center}
\includegraphics[width=1.6\columnwidth]{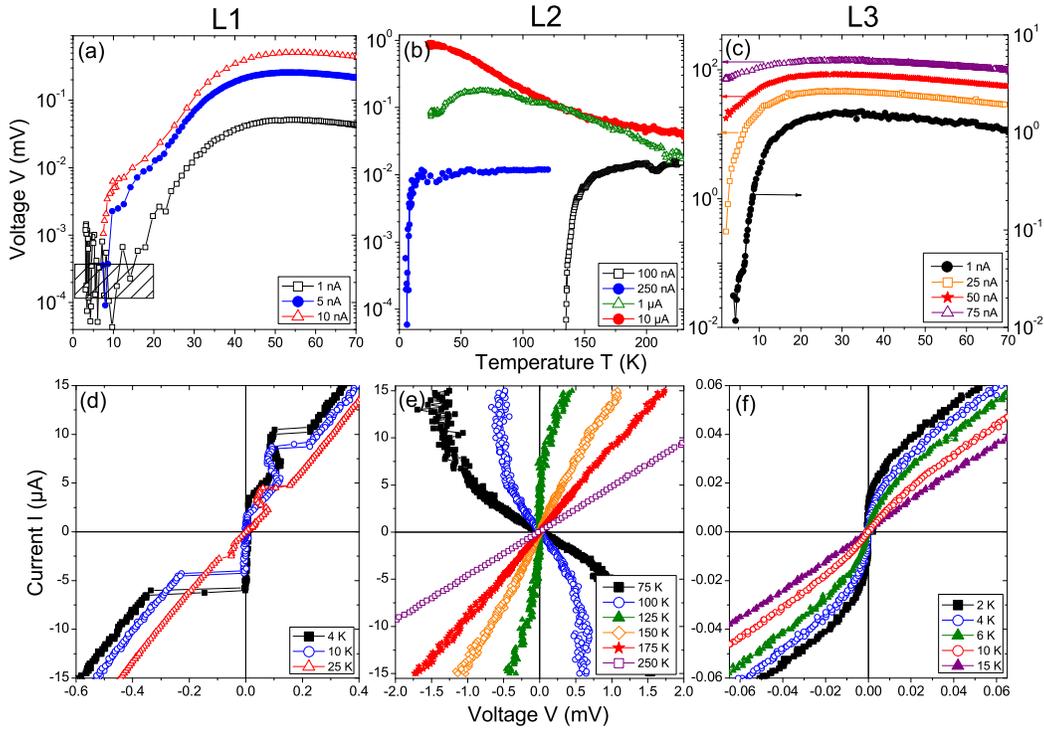}
\caption[]{(a-c): Voltage vs. temperature at different input currents for samples L1-L3.
The shadow region in (a) depicts the 300 nV noise region of sample L1 at low temperatures.
This noise is intrinsic of the sample and vanishes after the application of a magnetic field.
(d-f): $I-V$ characteristics curves at different temperatures for the same samples.} \label{Fig3}
\end{center}
\end{figure*}

 The observed $T-$dependence in Fig.~\ref{Fig2} is
determined mainly by the contribution of the internal interfaces,
since the graphene layers within the crystalline regions show a
semiconducting narrow-band-gap behavior\cite{gar12}. If the
voltage drop is related to some kind of granular superconductivity
we expect that its temperature dependence is sensitively
influenced by the applied current. Figures~\ref{Fig3} (a) to (c)
show clearly that the higher the input current the lower the
transition temperature, revealing a semiconducting-like behavior
at higher $I$ and $T$. The observed behavior is compatible with
the existence of granular superconductivity (see, e.g.,
Ref.~\onlinecite{ber86})  embedded in some of the internal
interfaces.

\begin{figure} []
\begin{center}
\includegraphics[width=1\columnwidth]{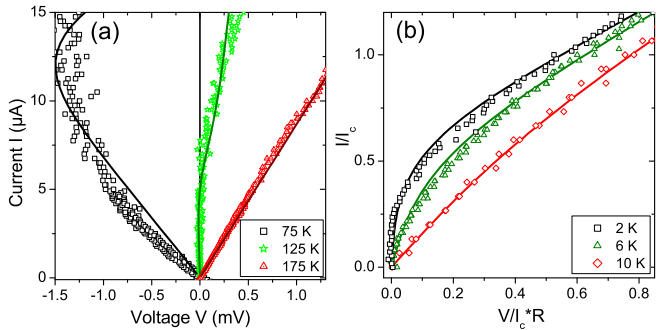}
\caption[]{Current-Voltage characteristics at different
temperatures for sample L2 (a) and L3 (b), this last in reduced
coordinates where $R$ is the normal state resistance. The
continuos curves are fits to the model proposed in
Ref.~\protect\onlinecite{amb69} with the Josephson critical
current $I_c(T)$ as the only free parameter.} \label{Fig4}
\end{center}
\end{figure}

Further support to this claim is obtained from the $I-V$ characteristics curves.
Samples L1 (Fig.~\ref{Fig3}(d)) and L3 (Fig.~\ref{Fig3}(f))
show typical Josephson behavior.  In the case of L2
(Fig.~\ref{Fig3}(e)) the curves were obtained with the
van der Pauw configuration and a current and voltage paths that  catch the
answer of more than one Josephson junction in that
sample. Taking into account the used current distribution
in this case and assuming two Josephson junctions with different $T_c$'s
the rather exotic
$I-V$ curves can be quantitatively understood using
the model of Ambegaokar and Halperin
\cite{amb69} where the influence of thermal fluctuations on
the dc Josephson effect in a junction of small capacitance
are taking into account, see Fig.~\ref{Fig4}(a). The same model can be successfully used to fit
the $I-V$ characteristics for sample L3 assuming only one Josephson junction,
see~\ref{Fig4}(b).  In the case of sample L1, a sharp jump
in the current appears at the corresponding critical Josephson current $I_c$
where also a small hysteresis is observed, see Fig.~\ref{Fig3}(d).

\begin{figure}[]
\begin{center}
\includegraphics[width=.7\columnwidth]{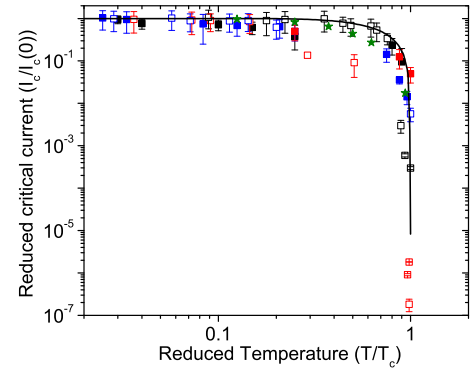}
\caption[]{Normalized Josephson critical current $I_c/I_c(0)$ vs.
normalized $T/T_c$ obtained for different lamellae (each symbol
corresponds to different a sample and/or configuration in the
measurements). Critical currents between 55 nA and $5.5 ~\mu$A and
critical temperatures from 15 K up to 175 K have been used. The
continuous line is the theoretical curve taken from
\cite{hag10} without any free parameter, assuming a short
junction length.} \label{Fig5}
\end{center}
\end{figure}

\begin{figure}[]
\begin{center}
\includegraphics[width=.8\columnwidth]{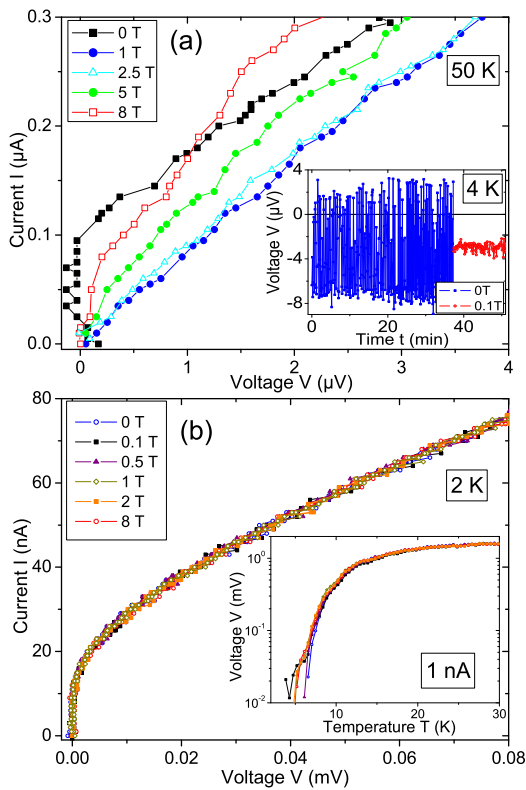}
\caption[]{(a) Current-Voltage curves for sample L4 at 50K with
and without magnetic field perpendicular to the graphene planes up
to 8T. The inset shows the time dependence of the voltage in
sample L2 for 250 nA input current at a constant temperature of 4
K, with no applied magnetic field (blue squares) and at 0.1~T
field applied perpendicular to the graphene planes (red dots). (b)
Current-Voltage characteristic curves for sample L3 at 2 K with
magnetic field applied parallel to the graphene planes up to 8 T.
The inset shows the temperature dependence of the voltage with
1~nA input current for sample L3 for magnetic fields from zero
(blue circles) up to 8~T (black squares) applied normal to the
graphene planes.} \label{Fig6}
\end{center}
\end{figure}

Considering the $I-V$ results from all samples and configurations
we obtain  $I_c(T)$ shown in
Fig.~\ref{Fig5} in normalized units. The overall behavior is
compatible with the temperature dependence expected for
Josephson-junctions where the normal barrier  is given by
ballistic graphene\cite{hag10}, the continuous line in Fig.~\ref{Fig5}.
One may ask whether a Josephson coupling is possible
through graphene layers and at large distances. Indeed,
the work in Ref.~\onlinecite{hee07} showed experimentally that
the Josephson effect is possible between superconducting
electrodes separated by hundreds of nanometers long graphene path.  Therefore,
 the assumption of  the existence of  superconducting regions
Josephson coupled through  graphene-like semiconducting paths at
the observed interfaces appears reasonable. Because these
superconducting regions at the interfaces are not homogeneously
distributed in the used samples, upon the junctions' distribution,
 a noisy or ``jumpy" behavior of the voltage is
 expected at low
 enough temperatures and currents due to phase  and current path fluctuations.
 We have  observed this behavior in samples L1, see Fig.~\ref{Fig3}(a), and L2,
see inset in Fig.~\ref{Fig6}(a). This noisy behavior is intrinsic
of the samples and can be suppressed by applying a magnetic field
normal to the graphene planes, see inset in Fig.~\ref{Fig6}(a), or
increasing $I$. Future work should reveal the frequency spectrum
of the noise and its relation to the Josephson junction
arrangement in the samples.

An applied magnetic field is expected to be detrimental to the
superconducting state. This effect can be due to an orbital
depairing effect, i.e. a critical increase of the shielding
currents, or at much higher fields  due to the alignment of the
electron spins, in case of singlet coupling. However, the possible
effects of a magnetic field on the superconducting state of quasi
two-dimensional superconductors or in case the coupling does not
correspond to a singlet state are not that clear. For example,
recently published experimental results\cite{gard11} in two
different two-dimensional superconductors, including one produced
at the interfaces between non superconducting regions,  show that
superconductivity can be even enhanced by a parallel magnetic
field. In case the pairing is $p-$type\cite{gon01} the influence
of a magnetic field is expected to be qualitatively different from
the conventional behavior\cite{sch80,kni98}  with even an
enhancement of the superconducting state at intermediate fields,
in case the orbital diamagnetism can be neglected or for parallel
field configuration. On the other hand, even for applied fields
normal to the planes we expect much less influence of the orbital
effect in case the London penetration depth is much larger than
the size of the superconducting regions at the interfaces of our
lamellae. If the superconducting coherence length is of the order
or larger than the thickness of the lamella then we expect a
superconducting (granular) behavior at lower $T$ but a magnetic
field may be less detrimental. We have studied the effects of
magnetic fields applied parallel and normal to the interfaces on
the transport characteristics in all measured lamellae. Upon
sample the observed effects are from an usual detrimental, no
effect at all or in some cases even a partial recovery of the
superconducting state.

Figure~\ref{Fig6} shows the $I-V$ characteristics of the samples
L4 (a) and L3 (b). For the relatively thick lamella L4 the
magnetic field of 1~T applied normal to the interface planes
vanishes the zero resistance state observed at zero field and at
50~K.  At higher fields, however, the $I-V$ curves show a recovery
to the zero resistance state. A field applied parallel to the
interface planes does not affect the curves. Recent studies on
possible superconductivity triggered by a large enough electric
field applied on multigraphene samples revealed also a kind of
reentrance above a certain magnetic field applied normal to the
graphene layers\cite{ana12}. For the thinnest sample L3, however,
a magnetic field applied in both directions has  little or no
effect on the $I-V$ characteristics or $V(T)$ up to 8~T, see
Fig.~\ref{Fig6}(b) and its inset.

In conclusion, the transport characteristics of thin lamellae with
tens of 2D interfaces between crystalline graphite regions reveal
Josephson-like behavior with zero resistance states at low enough
temperatures and input currents. Our results  finally clarify that
the origin for the metallic-like behavior as well as the giant
magnetic field induced metal-insulator (MIT) transition measured
in HOPG in the past\cite{kempa00,tok04,heb05} is related to the
superconducting properties of internal interfaces and it is not
intrinsic of the graphite structure. The existence of very high
temperature superconductivity embedded at these interfaces is
supported by recently done magnetization measurements\cite{she12}.


This work is supported by the Deutsche Forschungsgemeinschaft
under contract DFG ES 86/16-1. A.B. was supported by  ESF-Nano under the Graduate School of
Natural Sciences ``BuildMona".


%

\end{document}